\documentclass[aps,pre,twocolumn,groupedaddress,showpacs]{revtex4}

\usepackage{graphicx}

\begin{document}

\title{Fcc-bcc transition for Yukawa interactions determined by applied strain 
deformation}

\author{Robert S. Hoy}
\email{robhoy@pha.jhu.edu}
\author{Mark O. Robbins}

\affiliation{Department of Physics and Astronomy, Johns Hopkins University, 
Baltimore, Maryland 21218}

\date{March 5, 2004}

\begin{abstract}
Calculations of the work required to transform between bcc and fcc phases 
yield a high-precision bcc-fcc transition line for monodisperse point Yukawa 
(screened-Couloumb) systems.  Our results agree qualitatively but not 
quantitatively with recently published simulations and phenomenological 
criteria for the bcc-fcc transition.  In particular, the bcc-fcc-fluid triple 
point lies at a higher inverse screening length than previously reported.

\end{abstract}

\maketitle
\section{Introduction}
\label{sec:Intro}
The screened-Coulomb or Yukawa pair potential, $U(r) = \Phi e^{-\kappa r}/r$, 
has been the focus of great theoretical interest for two reasons. One is that 
it describes a wide range of interactions, changing continuously from a pure 
Coulomb potential to an effective hard-sphere potential as the inverse 
screening length $\kappa$ increases.  The second is that it provides an 
approximate description of the effective interactions between large ions that 
are screened by more mobile counterions.  In this context it has been used to 
describe the interactions between ions surrounded by electrons in metals 
\cite{Ashcroft}, dust grains surrounded by electrons in dusty plasmas 
\cite{Whipple, HF1994, Rosenfeld}, and macroions surrounded by counterions in 
charge-stabilized colloidal suspensions 
\cite{Rosenfeld, Verwey, Alexander, StevFalkRobb, Palberg}.

The phase diagram of systems of particles interacting with a Yukawa potential 
has been studied with both analytic 
\cite{Hone,Shih,Vaulina2002a,Vaulina2002b,Tejero,RT1987} 
and numerical \cite{RT1987, MF1991, ZKCA1992, SR1993, HF1994b, HFD1996,
KRG1986, RKG1988, DMRB1993, HFD1997, Reinhardt2000} techniques and compared 
to experiments on dusty plasmas \cite{ChuI, Nefedov2003} and colloidal 
suspensions \cite{Monovoukas, Sirota, STP1998}. 
The high temperature phase is a fluid.  There is no liquid-gas transition 
because the interactions are purely repulsive.  The stable crystalline 
phase at zero temperature changes from bcc to fcc as $\kappa$ increases.  The 
higher entropy of the bcc phase leads to a greater range of stability as 
temperature increases until the melting line is reached.  Previous results for 
the fcc-bcc transition line \cite{Hone, KRG1986, RKG1988, DMRB1993, HFD1997, 
Reinhardt2000} vary substantially and the most recent detailed calculation 
\cite{HFD1997} quotes an uncertainty of about 10\% roughly halfway between the 
zero-temperature transition point and the triple point. 

In this paper we use a different approach to obtain the bcc-fcc phase boundary 
with an uncertainty of only about 1\%.  Bounds on the free energy difference 
between the two phases are obtained by calculating the work done during a 
continuous deformation between them.  The effect of deformation rate, 
truncation of the potential, and system size and geometry are all analyzed to 
determine systematic errors.  The resulting bcc-fcc transition line is in 
qualitative agreement with recent simulation results, and quantitative 
differences are comparable to the larger error bars quoted by previous 
studies.  We estimate the location of the bcc-fcc-fluid triple point using 
previously published melting-line results \cite{SR1993,HFD1997}, and find 
that it lies at higher inverse screening lengths than previously reported.  

The role of anharmonicity in stabilizing the fcc phase is analyzed in detail. 
While anharmonicity increases the energy of the fcc phase relative to that of 
the bcc, there is an even larger increase in the relative entropic 
contribution to the free energy that increases the range of stability of 
the fcc phase.  This appears to reflect an increase in the frequency of the 
long wavelength shear modes that dominate the bcc entropy in the harmonic 
approximation \cite{RKG1988}.

Our results are also compared to phenomenological criteria proposed by Vaulina 
et.\ al.\ \cite{Vaulina2002a}. These authors predict a transition at a 
critical value of the mean-squared displacement about lattice sites, and 
calculate the displacement using a simple Einstein-like model.  We find that 
the actual displacement from MD simulations on our transition line is in 
reasonable agreement with their phenomenological criterion, but substantially 
larger than predicted by their Einstein model.  

The details of our calculations are presented in the following section.  
Section \ref{sec:result} provides a detailed analysis of systematic errors and 
presents our results for the phase boundary.  In Section \ref{sec:compare}, we 
compare our results to previous transition lines, and Section 
\ref{sec:conclude} provides a summary and conclusions.

\section{Method}
\subsection{Free energy difference calculations}
NVT ensembles are most natural for the study of Yukawa systems for two 
reasons.  First, since the Yukawa potential is purely repulsive, the 
macroions in an experiment will expand to fill the container.  Second, the 
inverse screening length $\kappa$ is density dependent in charged colloidal 
suspensions and dusty plasmas \footnote{$\kappa$ typically depends explicitly 
on the densities of the mobile counterions that screen interactions between 
macroions.  These densities change with macroion density to maintain charge 
neutrality.}.  This density dependence is system-specific, and affects the 
pressures and bulk moduli.  Thus any calculation of coexistence regions will 
be non-universal.  For this reason we focus on finding the Helmholtz free 
energy difference $\Delta F = F_{fcc} - F_{bcc}$ at fixed volume.  A brief 
discussion of coexistence is given in Section \ref{subsec:potpar}.

Postma, Reinhardt, and others \cite{Postma1986, Reinhardt1992, Reinhardt1993} 
have shown that the free energy difference between two phases of a system may 
be calculated in numerical simulations by evaluating the external work done 
on the system along a thermodynamic path connecting the phases.  From 
elementary thermodynamics, the mechanical work $W_{AB}$ done on a system on 
an isothermal path from state A to state B gives an upper bound on the change 
$\Delta F_{AB} = F_{B} - F_{A}$ in the system's Helmholtz free energy 
\cite{Reif}.  The work $W_{BA}$ done on the system during the reverse process 
$B\to A$ is an upper bound on $\Delta F_{BA}$, and hence $(-W_{BA})$ is a 
lower bound on $\Delta F_{AB}$.

\begin{figure}
\includegraphics[width=3.2in]{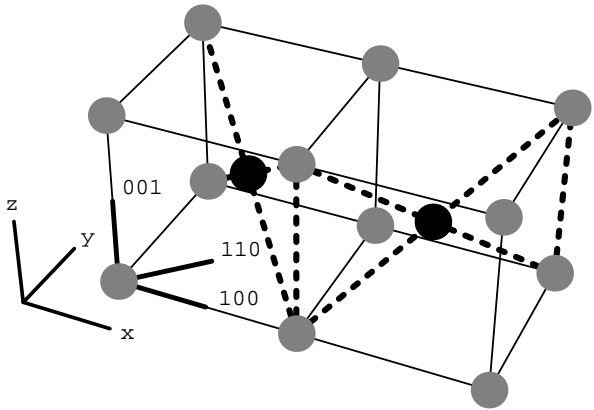}
\caption{The Bain transformation.  Two cells of a bcc lattice are shown with 
lattice directions.  The body-center atoms are shown in black.  When the x, y, 
and z directions are scaled by $(2^{-1/6},2^{-1/6},2^{1/3})$, the crystal is 
transformed into an fcc lattice of the same density.  The atoms connected by 
the dotted lines become two (100) faces of an fcc unit cell.}
\label{fig:baintrans}
\end{figure}

Bounds on $\Delta F$ for Yukawa systems can be obtained using a continuous 
constant-volume Bain deformation path (Fig.\ \ref{fig:baintrans}) connecting 
the bcc and fcc lattices.  An initially bcc lattice deformed such that its 
three cubic symmetry directions are scaled respectively by 
$(\zeta^{-1/2},\zeta^{-1/2},\zeta)$ is transformed into an fcc lattice of the 
same density as $\zeta$ varies continuously from 1 to $2^{1/3}$ 
\cite{Milstein1994}.  We calculate the work done along this path in the 
forward and reverse directions using strain-controlled molecular dynamics 
simulations \cite{Hoover1980}.

Assuming that the systems traverse these paths homogeneously, we can calculate 
the work done from the global stresses and strains.  We define 
\begin{equation}
W_{bf} = V\int_{bcc}^{fcc} \bar{\sigma}\cdot d\bar{\epsilon},
\label{eq:workstressstrainbf}
\end{equation}
\begin{equation}
W_{fb} = V\int_{fcc}^{bcc} \bar{\sigma}\cdot d\bar{\epsilon},
\label{eq:workstressstrainfb}
\end{equation}
where $\bar{\sigma}$ and $\bar{\epsilon}$ are the stress and true strain 
tensors.  $W_{bf}$ and $(-W_{fb})$ are upper and lower bounds on $\Delta F$.  
For these bounds to be narrow, the intermediate configurations of our systems 
must remain statistically representative of the $\zeta$-dependent equilibrium 
distributions as $\zeta$ is varied \cite{Postma1986}.  In particular, the 
stress tensor $\bar{\sigma}(\zeta)$ must remain near its equilibrium value.

\subsection{Potential parameters}
\label{subsec:potpar}
The phase behavior of Yukawa systems is most conveniently expressed in terms 
of dimensionless screening and temperature parameters.  One natural, phase-
independent length scale is $a = n^{-1/3}$, where $n = N/V$ is the macroion
number density.  The Yukawa potential may then be expressed as
\begin{equation}
U(r) = \frac{\Phi}{a}\frac{e^{-\lambda r/a}}{r/a},
\label{eq:scaledpot}
\end{equation}
where $\lambda = \kappa a$ is the dimensionless screening parameter.  The 
limits $\lambda\to0$ and $\lambda\to\infty$ correspond to the exhaustively 
studied one-component plasma and hard-sphere systems.  

A natural time scale is provided by $\tau_{E}$, the period of an Einstein 
oscillator in a crystal.  The Einstein periods for the fcc and bcc phases 
change by an order of magnitude over the range of $\lambda$ studied here 
$(3 \leq \lambda \leq 8)$, yet differ from each other by less than 1.2\% at 
any given $\lambda$ within this range.  To obtain consistent results across a 
wide range of screening lengths, we normalize all time scales in this study to 
$\tau_{E}(\lambda)$, using the fcc values given in Ref. \cite{RKG1988}.

A natural energy scale is given by the Einstein phonon energies 
$m\omega_{E}^{2}a^{2}$, where m is the macroion mass and $\omega_{E} = 
2\pi/\tau_{E}$ is the Einstein frequency.  Following Kremer, Robbins, and 
Grest \cite{KRG1986}, we define the dimensionless temperature
\begin{equation}
\widetilde{T} =  \displaystyle\frac{k_{B}T}{m\omega^{2}_{E}a^{2}}
\label{eq:Ttilde}
\end{equation}
using the fcc phonon energies, and plot our phase diagram in $(\lambda, 
\widetilde{T})$ space.  A dimensionless inverse temperature $\Gamma = 
(\Phi/a)/k_{B}T$ called the coupling parameter is used in many studies of 
dusty plasmas.  The advantage of using $\widetilde{T}$ rather than $\Gamma$ in 
Yukawa phase diagrams is that the transition lines are approximately linear in 
$\lambda$. 

The bcc and fcc phases coexist in equilibrium over a part of the phase 
diagram.  Following previous authors \cite{KRG1986,RKG1988,DMRB1993,HFD1997}, 
we define the bcc-fcc transition line as the curve 
$\widetilde{T}_{trans}(\lambda)$ on which 
$\Delta F = \Delta F(\lambda, \widetilde{T}) = 0$.  This transition line 
will certainly lie within the coexistence region, regardless of the 
thermodynamic state dependence of $\kappa$ and $\Phi$.  We find 
$\widetilde{T}_{trans}(\lambda)$ by calculating $\Delta F$ at many points 
$(\lambda_{i}, \widetilde{T}_{i})$ on the phase diagram.

\subsection{MD simulation details}
\label{subsec:MDdetails}
We simulate NVT ensembles of identical particles using a velocity-Verlet 
\cite{VelocVerlet} algorithm to integrate the particle trajectories.  The 
temperature is maintained with a Langevin thermostat \cite{Schneider}.  
Periodic boundary conditions are used to maintain the density.  The equations 
of motion for the position $\vec{q}_{i}$ and peculiar momentum $\vec{p}_{i}$ 
of the \textit{i}th particle are
\begin{equation}
\begin{array}{ccl}
\dot{\vec{q}}_{i} &=& \vec{p}_{i}/m + \dot{\bar{\epsilon}}\vec{q}_{i},\\
\dot{\vec{p}}_{i} &=& \vec{F}_{i} - \dot{\bar{\epsilon}}\vec{p}_{i} + \vec{\eta}_{i} - \vec{p}_{i}/\tau_{Lang},
\end{array}
\label{eq:eqnsofmotion}
\end{equation}
where $\dot{\bar{\epsilon}}$ is the true strain rate tensor, $\vec{F}_{i}$ is 
the force due to Yukawa interactions, $\vec{\eta}_{i}$ is a random noise term, 
and $\tau_{Lang}$ is the characteristic relaxation time of the thermostat.  We 
use a timestep $\delta t = .01\tau_{E}$ to insure proper integration of Eqs.\ 
(\ref{eq:eqnsofmotion}) and set $\tau_{Lang} = 10\tau_{E}$.  Changing 
$\delta t$ and $\tau_{Lang}$ by a factor of two in either direction had no 
effect on the phase diagram.

For numerical efficiency we truncate interactions at a cutoff radius $r_{c}$.  
Due to the presence of long range order in Yukawa crystals, care must be taken 
in choosing this cutoff radius.  We present the details of our determination 
of $r_{c}(\lambda)$ in Section \ref{subsec:pcd}.

In most of our simulations, we impose the bcc$\to$fcc Bain transformation as 
follows.  We start with a lattice of 3456 particles ($12^{3}$ bcc unit cells) 
in a cubic simulation cell with edges of length $L_{x}=L_{y}=L_{z}=L_{0}$ 
aligned with the $<100>$ directions of the lattice.  The system is 
equilibrated for 200 Einstein periods.  We then fix 
$\dot{\zeta} = \dot{L_{z}}/L_{0}$ for a time $\Delta t$ sufficient to reach 
the fcc structure: $\Delta t = (2^{\frac{1}{3}} - 1)\dot{\zeta}^{-1}$.  
The other cell edges $L_{x}$ and $L_{y}$ are varied to maintain constant 
volume and tetragonality $(L_{x}=L_{y}=\sqrt{L_{0}^{3}/L_{z}})$.  The true 
strain rate tensor $\dot{\bar{\epsilon}}$ is then given by
\begin{equation}
\begin{array}{l}
\dot{\epsilon}_{zz} = \dot{L_{z}}/L_{z} = \dot{\zeta}/\zeta,\\
\dot{\epsilon}_{xx} = \dot{\epsilon}_{yy} = \dot{L_{x}}/L_{x} = -\dot{\zeta}/2\zeta,\\
\dot{\epsilon}_{xy} = \dot{\epsilon}_{xz} = \dot{\epsilon}_{yz} = 0.
\end{array}
\label{eq:strainratetensor}
\end{equation}

We compute the diagonal elements $(P_{x}, P_{y}, P_{z})$ of the pressure 
tensor using standard methods \cite{Allen}.  Equation 
(\ref{eq:workstressstrainbf}) then takes on the more physically familiar form
\begin{equation}
W_{bf} = -\small\int_{0}^{\Delta t}\left(P_{x}L_{y}L_{z}\dot{L}_{x} +  
P_{y}L_{x}L_{z}\dot{L}_{y} +  P_{z}L_{x}L_{y}\dot{L}_{z}\right)dt.
\label{eq:workfrompres}
\end{equation}
After the system has reached the fcc structure, the deformation process is 
reversed by changing the sign of $\dot{\zeta}$.  As the system returns to bcc, 
$W_{fb}$ is calculated using the analogue of Eq.\ (\ref{eq:workfrompres}).  

To minimize uncertainties in $\widetilde{T}_{trans}(\lambda)$, $\dot{\zeta}$ 
must be small enough for the system to remain near equilibrium.  One 
requirement is that the strain-rate components 
$(\dot{\bar{\epsilon}}\vec{q}_{i})$ of the velocities must be small compared 
to the thermal velocity.    The Bain transformation time $\Delta t$ (which is 
proportional to $\dot{\zeta}^{-1}$) must also be large compared to 
$\tau_{Lang}$ to allow the thermostat to transfer heat to or away from the 
system as necessary to maintain constant temperature.  Since $\tau_{Lang}$ 
sets the time over which the system samples the canonical ensemble,  the 
thermodynamic sampling improves as  $\Delta t/\tau_{Lang}$ increases. 

The precision of the calculated transition line depends on the difference 
$(\Delta F_{max} - \Delta F_{min}) \equiv W_{bf} + W_{fb} = W_{cycle}$ between 
the bounds on $\Delta F$. These bounds converge to each other in the 
reversible thermodynamic (zero strain rate) limit. In this limit, the average 
work $<W_{cycle}>$ done on the system over a full deformation cycle 
(bcc$\to$fcc$\to$bcc or vice versa) should vanish.  In simulations at finite 
strain rate, however, there is a positive systematic error in $W_{cycle}$ due 
to energy dissipation \cite{Jarzynski}.  
This can be physically interpreted as arising from viscosity.  Each applied 
strain increment takes the system slightly out of equilibrium.  When 
$\dot{\zeta}$ is small, one expects the stresses to deviate from their 
equilibrium values by an amount $\bar{\sigma}_{visc} \sim \dot{\zeta}$ 
\cite{Larson}.  Sources of viscous dissipation include the intrinsic viscosity 
and the drag forces $-\vec{p}_{i}/\tau_{Lang}$ on the particles applied by the 
Langevin thermostat.  The viscous dissipation rate is given by 
$\bar{\sigma}_{visc}\cdot\dot{\bar{\epsilon}}$, so one expects the dissipated 
power to be proportional to $\dot{\zeta}^{2}$.  Since the total simulation 
time scales as $\dot{\zeta}^{-1}$, the total dissipated energy, and hence the 
deviation of $<W_{cycle}>$ from zero, should be linearly proportional to 
$\dot{\zeta}$.  We present the $\dot{\zeta}$-dependence of our results in 
Section \ref{subsec:srd}.  

\section{Results}
\label{sec:result}
\subsection{Strain rate dependence}
\label{subsec:srd}
At temperatures near the transition line, calculations of the geometrical 
structure factor and pair correlation function verify that our systems 
traverse the Bain transformations homogeneously.  This homogeneity allows us
to use Eqs.\ (\ref{eq:workstressstrainbf},\ref{eq:workstressstrainfb})
for calculating 
$W_{bf}$ and  $W_{fb}$ and leads to tight bounds on $\Delta F$. 

\begin{figure}[h]
\includegraphics[width=3.2in]{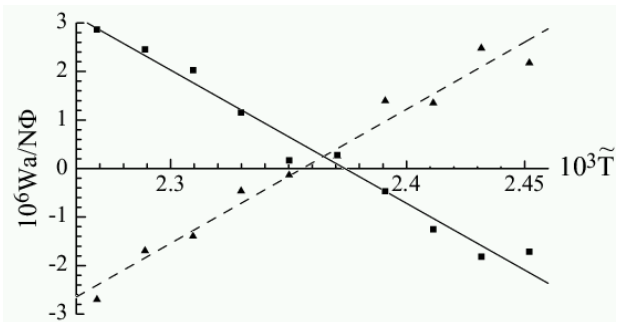}
\caption{Simulation results for $\lambda = 5, 
\dot{\zeta} = 10^{-4}/\tau_{E}$.  Solid triangles are values of $W_{bf}$, 
solid squares are values of $W_{fb}$.  The dashed and solid lines are linear 
fits to the results.  $\widetilde{T}_{bf}$ and $\widetilde{T}_{fb}$ are the 
intersections of these lines with W = 0.}
\label{fig:lam5work}
\end{figure}

We obtained results similar to those shown in Figure \ref{fig:lam5work} over 
the entire investigated range of $\lambda$ and for three different values of 
$\dot{\zeta}$.  Near the transition line, $W_{bf}(\lambda, \widetilde{T})$ and 
$W_{fb}(\lambda,\widetilde{T})$ vary linearly with $\widetilde{T}$ and have 
nearly opposite ($\lambda$-dependent) slopes.  The scatter about linear fits 
to $W_{bf}(\lambda, \widetilde{T})$ and $W_{fb}(\lambda,\widetilde{T})$ is 
consistent with fluctuations in $W_{bf}$ and $W_{fb}$ at fixed $(\lambda, 
\widetilde{T})$.    The intersections of these fits with $W = 0$ give two 
estimates, $\widetilde{T}_{bf}$ and $\widetilde{T}_{fb}$, for 
$\widetilde{T}_{trans}$.  These are obtained using data at ten evenly spaced 
$\widetilde{T}$ within about 5\% of the transition line.

\begin{figure}
\includegraphics[width=3.2in]{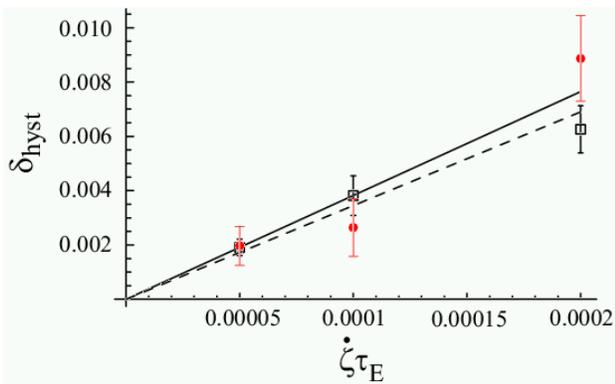}
\caption{Strain-rate dependence of $\delta_{hyst}$.  Solid circles indicate 
$\lambda = 4$ results.  Empty squares indicate $\lambda = 5$ results.  The 
solid and dashed lines are linear fits to the data.  The error bars indicate 
statistical uncertainties. (Color online)}
\label{fig:deltahyst}
\end{figure}

For a given system, $\widetilde{T}_{fb}$ and $\widetilde{T}_{bf}$ provide 
upper and lower bounds on $\widetilde{T}_{trans}$ since 
$W_{bf} > \Delta F > -W_{fb}$ and 
$(\partial{\Delta F}/\partial{\widetilde{T}}) > 0$.  We define the fractional 
uncertainty due to dissipative hysteresis as
\begin{equation}
\delta_{hyst} = (\widetilde{T}_{fb} - \widetilde{T}_{bf})/(\widetilde{T}_{fb} 
+ \widetilde{T}_{bf}).
\label{eq:deltahyst}
\end{equation}
The conditions $\delta_{hyst} \sim \dot{\zeta}$ and $<W_{cycle}> \sim 
\dot{\zeta}$ are equivalent due to the linear dependence of $W_{bf}$ and 
$W_{fb}$ on $\widetilde{T}$.  Figure \ref{fig:deltahyst} shows the 
$\dot{\zeta}$-dependence of $\delta_{hyst}$ for $\lambda = 4$ and 
$\lambda = 5$.  The results are consistent with our hypothesis that the 
energy dissipated is linear in $\dot{\zeta}$.  

We identify $\widetilde{T}_{trans} = (\widetilde{T}_{bf} + 
\widetilde{T}_{fb})/2$ as the best estimate for the transition temperature 
for a given system size, system geometry, and potential cutoff radius.  Table 
\ref{tab:strainratedep} shows that the fractional variation of 
$\widetilde{T}_{trans}$ with $\dot{\zeta}$ is much smaller than 
$\delta_{hyst}$ \footnote{These values are for the standard system size and 
geometry described in Section \ref{subsec:MDdetails} and cutoff radii 
reported in Section \ref{subsec:pcd}.}.     

\begin{table}[h]
\caption{Dependence of $\widetilde{T}_{trans}(\lambda)$ on dimensionless 
strain rate.}
\begin{ruledtabular}
\begin{tabular}{lll}
$\dot{\zeta}\tau_{E}$ &  $10^{3}\widetilde{T}_{trans}(\lambda = 4)$ &  
$10^{3}\widetilde{T}_{trans}(\lambda = 5)$\\
$5\cdot 10^{-5}$ & $1.637\pm.003$ & $2.363\pm.002$\\
$1\cdot 10^{-4}$ & $1.634\pm.004$ & $2.365\pm.004$\\
$2\cdot 10^{-4}$ & $1.633\pm.005$ & $2.366\pm.005$\\
\end{tabular}
\end{ruledtabular}
\label{tab:strainratedep}
\end{table}

In the following, we present results for $|\dot{\zeta}|=10^{-4}/\tau_{E}$.  
Based on  Table \ref{tab:strainratedep}, for this value of $|\dot{\zeta}|$ the 
random and finite strain rate uncertainties in $\widetilde{T}_{trans}$  are 
comparable, both about 0.2\%.  The combined error is estimated to be less than 
0.4\%.  The uncertainties given in subsequent tables include only statistical 
uncertainties from the linear fits used to calculate $\widetilde{T}_{bf}$ and 
$\widetilde{T}_{fb}$.

\subsection{Potential cutoff dependence}
\label{subsec:pcd}
We estimate the errors introduced by truncating the force at $r_{c}$ by 
calculating the error $\delta E(r_{c})$ in the potential energy difference.  
If the error in $\Delta F$ is of the same order, then the fractional error in 
the transition temperature is
\begin{equation}
\delta_{cut} \equiv \frac{\delta \widetilde{T}(r_{c})}{\widetilde{T}} \simeq 
\frac{1}{\widetilde{T}} \frac{\partial{\widetilde{T}}}{\partial \Delta F}\delta E(r_{c}).
\label{eq:deltacut}
\end{equation}
Here $(\partial \widetilde{T}/\partial \Delta F)$ is known near the transition 
line from the work calculations.  

The cutoff-induced error in the potential energy difference can be written in 
terms of the pair correlation functions $g_{fcc}(r)$ and $g_{bcc}(r)$ of the 
fcc and bcc crystals.  For N particles 
\begin{equation}
\delta E(r_{c}) = \frac{N}{2}\int_{r_{c}}^{\infty} U(r)(g_{fcc}(r) - 
g_{bcc}(r))4\pi r^{2}dr/a^3
\label{eq:exactcutint}
\end{equation}
If $\Omega \equiv \textrm{max}\{|g_{fcc}(r) - g_{bcc}(r)|; r \geq r_{c}\}$, 
then
\begin{equation}
\small\frac{|\delta E(r_{c})|}{N(\Phi/a)} < \frac{\Omega}{\Phi a^2}\int_{r_{c}}^{\infty} U(r) 2\pi r^{2}dr = \frac{2\pi\Omega (1+\lambda r_{c}/a)e^{-\lambda 
r_{c}/a}}{\lambda^2}
\label{eq:ubcutint}
\end{equation}
One expects $\Omega$ to be of order 1 at finite temperature. 

For $\lambda$ = 3, 4, and 5, we estimated $\delta E(r_{c})$ at $\widetilde{T} 
\simeq \widetilde{T}_{trans}$ by calculating the pair correlation functions in 
large systems using large cutoff radii and long integration times.
Due to the exponential falloff of U(r) and finite temperature smoothing of 
g(r), the infinite upper bound in Eq.\  \ref{eq:exactcutint}) can be replaced 
by a finite value $r_{l}$ without introducing significant errors.  We found 
$\lambda r_{l} = 30a$ to be sufficiently large. 
 
\begin{figure}
\includegraphics[width=3.2in]{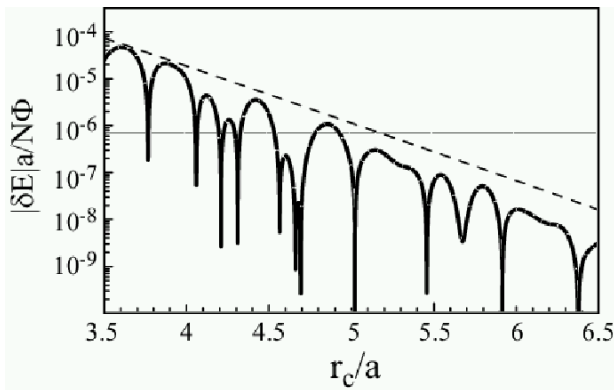}
\caption{Determination of $r_{c}$ for $\lambda = 3$.  The heavy curve is our 
estimate of $\delta E(r_{c})$ obtained from simulations at $\widetilde{T} = 
8.84\cdot10^{-3}$, with $r_{l} = 10a$.  The dashed line is the analytic upper 
bound on $\delta E(r_{c})$ from Eq.(\ref{eq:ubcutint}) with $\Omega = 1/3$.    
The horizontal line is the value of $|\delta E|$ in Eq.(\ref{eq:deltacut}) 
corresponding to $|\delta_{cut}| = 0.01$.}
\label{fig:deltaE}
\end{figure}

Figure \ref{fig:deltaE} shows our estimate of $|\delta E(r_{c})/(N\Phi/a)|$ 
from Eq.\ (\ref{eq:exactcutint}) and the value of $|\delta E/(N\Phi/a)|$ 
corresponding to $\delta_{cut} = 0.01$ for $\lambda = 3$, the longest-range 
potential considered.  We found that $\Omega$ decreases from 1.4 to .62 as 
$r_{c}$ increases from 3.5a to 6.5a.  The actual error is always smaller than 
the bound given by $\Omega$ because $g_{fcc} - g_{bcc}$ oscillates in sign.  
The envelope shown corresponds to $\Omega = 1/3$.  Because of the sharp 
variation of $\delta E(r_{c})$, we use the envelope to estimate 
$\delta_{cut}$.  

To test Eq. (\ref{eq:deltacut}) we calculated $\widetilde{T}_{trans}$ as a 
function of $r_{c}$ for $\lambda = 3$.  Results are shown in Table 
\ref{tab:cutoffdep}.  The fractional changes in $\widetilde{T}_{trans}$ from 
$r_{c} = 3.5a$ and $r_{c} = 4.667a$ to $r_{c} = 6.667a$ are 19\% and 0.9\%, 
respectively.  Both changes are about one-fifth of the estimates for 
$\delta_{cut}$ from Eqs.\ (\ref{eq:deltacut},\ref{eq:exactcutint}).  No 
statistically significant changes are expected or observed for $r_{c} \geq 
5.833a$.  We conclude that errors estimated from the envelopes of curves like 
Fig. \ref{fig:deltaE} give a conservative estimate of cutoff errors.  

\begin{table}[h]
\caption{$\widetilde{T}_{trans}$ vs. $r_{c}$ for $\lambda = 3$. 
$|\delta_{cut}|$ is given by Eq.\ \ref{eq:deltacut} and the bound in
Eq.\ \ref{eq:ubcutint}.}
\begin{ruledtabular}
\begin{tabular}{lll}
$r_{c}$ & $|\delta_{cut}|$ & $10^{3}\widetilde{T}_{trans}$\\
3.5a & 1.04 & $1.048\pm.003$\\
4.667a & $0.041$ & $0.871\pm.002$\\
5.833a & $0.0015$ & $0.880\pm.002$\\
6.667a & $0.00014$ & $0.879\pm.003$\\
\end{tabular}
\end{ruledtabular}
\label{tab:cutoffdep}
\end{table}

To ensure that the fractional systematic errors were no larger than our random 
and rate errors, we chose $r_{c}$ slightly above the values corresponding to 
$|\delta_{cut}| = 0.002$.  For $\lambda$ = 3, 4, and 5, we used cutoff radii 
of 5.833a, 4.375a, and 3.5a in the simulations used to determine 
$\widetilde{T}_{trans}$.  Smaller $r_{c}$ can be used at higher $\lambda$ both 
because the interactions weaken and 
$\widetilde{T}_{trans}/\widetilde{T}_{melt}$ increases, leading to a smaller 
$\Omega$.  For $\lambda \geq 5$ we fixed the cutoff radius at $r_{c} = 3.5a$.  

\subsection{System size and geometry dependence}
To examine finite size effects we also considered a 432-particle system 
(initial state $6^{3}$ bcc unit cells).  Because the corresponding fcc state 
has transverse length $6.73a$, the minimum image convention requires 
$r_{c} < 3.367a$, and we used $r_{c} = 3.3a$.  To separate out 
$r_{c}$-dependence from system size dependence, we also recalculated the 
transition line for $N = 3456$ for $5 \leq \lambda \leq 8$ with 
$r_{c} = 3.3a$.  

Table \ref{tab:sizedep} shows a comparison of our calculated transition 
temperatures.  The $N = 432$ values were systematically lower, but the effect 
was small.  From theoretical considerations one expects the leading finite 
size corrections to $\Delta F/N$ to be proportional to $1/N$ \cite{Polson}.  
This should produce a corresponding error in $\widetilde{T}_{trans}$.   As 
shown in Table \ref{tab:sizedep}, the changes in $\widetilde{T}_{trans}$ from 
$N = 432$ to $N=3456$ were all about 1\%.  The changes in 
$\widetilde{T}_{trans}$ from $N = 3456$ to $N = \infty$ for this system 
geometry should be about 8 times smaller.  

\begin{table}[h]
\caption{Dependence of $\widetilde{T}_{trans}$ on N.}
\begin{ruledtabular}
\begin{tabular}{lll}
$\lambda$ &  $10^{3}\widetilde{T}_{trans}(N = 432)$ &  
$10^{3}\widetilde{T}_{trans}(N = 3456)$\\
5 & $2.350\pm.009$ & $2.362\pm.004$\\
6 & $2.985\pm.011$ & $3.021\pm.003$\\
7 & $3.562\pm.009$ & $3.593\pm.005$\\
8 & $4.064\pm.008$ & $4.087\pm.005$\\
\end{tabular}
\end{ruledtabular}
\label{tab:sizedep}
\end{table}

Another test indicates that finite size effects are larger than the above 
estimate.  The geometry was changed so that the fcc state has equal cell edges 
and the bcc state has $L_{x} = L_{y} = \sqrt{2}L_{z}$.  These simulations 
contained $10^{3}$ fcc unit cells (4000 particles) with the $<100>$ directions 
parallel to the simulation cell edges.  After Bain transformation, the bcc 
state has two $<110>$ directions parallel to the simulation cell edges.  As 
shown in Table \ref{tab:orientdep}, the values of $\widetilde{T}_{trans}$ 
obtained for both $\lambda = 4$ and $\lambda = 7$ were 0.6\% lower than those 
obtained with the standard system geometry \footnote{Both sets of simulations 
used the cutoff radii indicated in Section \ref{subsec:pcd}.}.  Other 
simulations verified that this was due solely to the change in boundary 
conditions.  We conclude that our dominant source of uncertainty is finite 
size and is less than 1\%.

\begin{table}[h]
\caption{Dependence of $\widetilde{T}_{trans}$ on system geometry.}
\begin{ruledtabular}
\begin{tabular}{lll}
$\lambda$ &  $10^{3}\widetilde{T}_{trans}(N = 3456)$ &  
$10^{3}\widetilde{T}_{trans}(N = 4000)$\\
4 & $1.634\pm.004$ & $1.625\pm.004$ \\
7 & $3.592\pm.004$ & $3.570\pm.004$\\
\end{tabular}
\end{ruledtabular}
\label{tab:orientdep}
\end{table}

We attribute the observed sensitivity to geometry to the change in allowed low 
frequency modes.  These modes play a disproportionate role in determining the 
entropy in lattice dynamics calculations \cite{RGK1990} and drive the 
fcc$\to$bcc transition with increasing temperature \cite{RKG1988}.  Since the 
shear velocity is highly anisotropic in the bcc phase, changing the boundaries 
affects the sampling of these low frequency modes and thus $\Delta F$.  

\subsection{Transition line}
Table \ref{tab:Ttrans} shows our calculated $\widetilde{T}_{trans}(\lambda)$ 
with statistical uncertainties.  As described above, the combined systematic 
errors due to finite strain rate, system size, and potential cutoff are 
estimated to be less than 1\%. Results of a cubic polynomial fit to the data 
are also given:
\begin{equation}
\begin{array}{lcl}
10^{4}\widetilde{T}_{trans}^{fit}(\lambda) &=&  6.46678(\lambda-\lambda_{0}) + 
0.43001(\lambda-\lambda_{0})^{2}\\
&&  - 0.06806(\lambda-\lambda_{0})^{3},
\end{array}
\label{eq:cubicTtransfit}
\end{equation}
where $\lambda_{0} = 1.718$ is the zero-temperature transition point obtained 
from lattice statics calculations \cite{Medeiros}.  Lower-order polynomials 
fail to adequately fit the data within our uncertainties.  

\begin{table}[h]
\caption{Calculated and fit values of $\widetilde{T}_{trans}$. Only 
statistical uncertainties are quoted.}
\begin{ruledtabular}
\begin{tabular}{lll}
$\lambda$ & $10^{3}\widetilde{T}_{trans}$ & 
$10^{3}\widetilde{T}^{fit}_{trans}$\\
3 & $0.880\pm.002$ & 0.885\\
4 & $1.634\pm.004$ & 1.619\\
5 & $2.365\pm.004$ & 2.345\\
6 & $3.017\pm.004$ & 3.023\\
7 & $3.592\pm.004$ & 3.613\\
8 & $4.085\pm.004$ & 4.072\\
\end{tabular}
\end{ruledtabular}
\label{tab:Ttrans}
\end{table}

\begin{figure}
\includegraphics[width=3.2in]{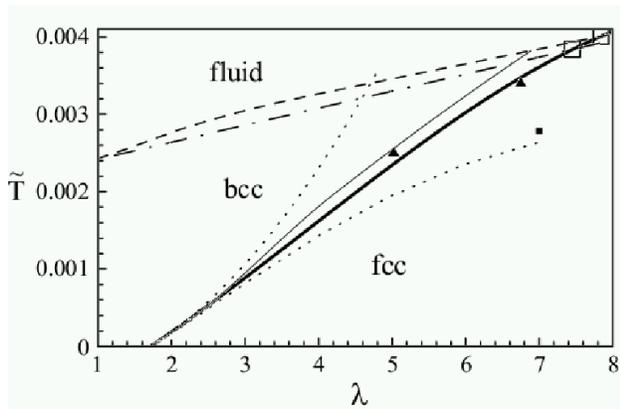}
\caption{Phase diagram of Yukawa systems.  The heavy line is our cubic 
polynomial fit for the bcc-fcc transition line.  The light solid line is the 
fcc-bcc transition line from Ref. \protect\cite{HFD1997}. The low and high 
dotted lines are respectively the lattice dynamics and molecular dynamics 
bcc-fcc transition lines from Ref.\ \protect\cite{RKG1988}.  The triangles 
are a bcc-fcc coexistence point and triple point from Ref.\ 
\protect\cite{DMRB1993}, and the solid square is a bcc-fcc transition point 
from Ref.\ \protect\cite{Reinhardt2000}. The dashed and dash-dotted lines are 
the melting lines from Ref.\ \protect\cite{SR1993} and Ref.\ 
\protect\cite{HFD1997}.   The empty squares denote our estimates for the 
bcc-fcc-fluid triple point.}
\label{fig:phasediag}
\end{figure}

Figure \ref{fig:phasediag} shows the polynomial fit and two previously 
published solid-fluid coexistence lines \cite{SR1993,HFD1997}.  The 
intersections of these lines give estimated values of the bcc-fcc-fluid 
triple point.  Using results from Ref.\ \cite{SR1993}  we find 
$(\lambda_{tp} = 7.45, \widetilde{T}_{tp} = 0.00384)$.  Those from 
Ref.\ \cite{HFD1997} yield 
$(\lambda_{tp} = 7.84, \widetilde{T}_{tp} = 0.00401)$.

If we assume that the parameters $\kappa$ and $\Phi$ are density-independent,
we can calculate the width of the bcc-fcc coexistence region from the 
pressures and bulk moduli of the two phases on the line where $\Delta F = 0$.
The bcc pressure is larger than the fcc pressure by only about 0.04\% for
$\lambda=4$, and by 0.65\% for $\lambda=7$.  This results in a higher 
density in the fcc phase at coexistence, but only by about 0.015\% at 
$\lambda=4$ and 0.2\% at $\lambda=7$.  The corresponding changes in $\lambda$ 
and $\widetilde{T}$ are much smaller than the uncertainties in our calculated 
transition line.  The coexistence region in experimental systems may be much 
larger due to variations in $\kappa$ and $\Phi$ with density 
\cite{STP1998,Nefedov2003}.  As noted above, these variations are system 
specific and a more complete treatment is beyond the scope of this paper.

\subsection{Anharmonic effects}
The dotted line in Figure \ref{fig:phasediag} shows lattice dynamics results 
for the bcc-fcc transition line \cite{RKG1988}.  In this approximation the 
energy and entropy differences, $\Delta E_{LD}$ and $\Delta S_{LD}$, are 
independent of T.  The fcc-bcc transition line is given by 
$T_{LD} = \Delta E_{LD}/\Delta S_{LD}$.  The resulting curve lies below 
$\widetilde{T}_{trans}$, indicating that the fcc phase is stabilized by 
anharmonic effects \cite{RKG1988,HFD1997}.  This implies that the anharmonic 
component of the free energy difference,
\begin{equation}
\Delta F_{an} \equiv \Delta E_{an} - T\Delta S_{an} = \Delta F - \Delta F_{LD},
\label{eq:deltaFanharm}
\end{equation}
is negative on the transition line.  The relative signs and magnitudes of 
$\Delta E_{an}$ and $\Delta S_{an}$ may be calculated by comparing our 
accurate measurements of free and total energy differences with the 
lattice-dynamics results. 

Table \ref{tab:anharmdeltaEFtab} shows results for anharmonic contributions to 
the free and total energy differences on the fit transition line 
\footnote{Results for $\lambda = 8$ are not presented in Tables 
\ref{tab:anharmdeltaEFtab} and \ref{tab:rmsd} respectively because at 
$\widetilde{T}_{trans}$ solid diffusion prohibits accurate calculations of the 
rms displacements, and lattice dynamics calculations of $\Delta S$ are 
unavailable.}.  The values of $\Delta F_{an}$ are known from the work 
calculations, while the values of $\Delta E_{an}$ were found from separate 
equilibrium simulations.  The anharmonic corrections to the total energy favor 
the bcc phase for all $\lambda$, i.e., $E_{fcc} - E_{bcc}$ on the transition
line has increased relative to its zero-temperature value.
The anharmonic contributions to the free energy difference, however, are 
larger in magnitude and opposite in sign, implying that anharmonic entropic 
contributions to $\Delta F$ favor the fcc phase at all $\lambda$ and overwhelm
energetic contributions.

\begin{table}[h]
\caption{Anharmonic free and total energy differences evaluated at 
$\widetilde{T}^{fit}_{trans}(\lambda)$.}
\begin{ruledtabular}
\begin{tabular}{lll}
$\lambda$ & $10^{2}\Delta F_{an}/Nk_{B}T$ & $10^{2}\Delta E_{an}/Nk_{B}T$\\
3 & $-0.58 \pm 0.1$ & $0.42 \pm 0.33$\\ 
4 & $-1.07 \pm .12$ & $1.0 \pm .4$\\ 
5 & $-1.92 \pm .15$ & $1.2 \pm .4$\\ 
6 & $-3.03 \pm .17$ & $2.4 \pm .4$\\ 
7 & $-4.33 \pm .21$ & $2.2 \pm .4$\\ 
\end{tabular}
\end{ruledtabular}
\label{tab:anharmdeltaEFtab}
\end{table}

In lattice calculations the larger entropy of the bcc phase comes mainly from 
the lower frequency of its shear modes. Some of these modes have negative 
energy for $\lambda > 7.67$, causing the bcc phase to become linearly unstable
at low temperatures \cite{RKG1988}.  It is interesting that the onset of this 
low temperature instability is close to $\lambda_{tp}$.  However, we have 
performed runs near the melting line for $\lambda$ as large as 10 and find 
that the bcc phase remains metastable.  This implies that anharmonic effects 
have increased the frequency of long wavelength shear modes, providing an 
explanation for the decreased entropy advantage of the bcc phase.

\section{Comparison to Previous Results}
\label{sec:compare}
Miller and Reinhardt were the first authors to use Bain deformation paths to 
obtain bounds on $\Delta F$ for Yukawa systems \cite{Reinhardt2000}.  They 
calculated the work by integrating the change in the Hamiltonian rather than 
from the stresses and strains.  The large discrepancy between their 
$\lambda = 7$ transition temperature and our result is likely due to their 
extremely small system size $(N = 108)$, which was just used to illustrate 
their method.

The earliest MD calculations \cite{KRG1986,RKG1988} of the transition line 
also deviate substantially from ours, particularly at large $\lambda$.  The 
line shown in Figure \ref{fig:phasediag} is a fit between points where the 
fcc and bcc phases were found to be stable.  The gap between points was about 
20\% and the final shape was strongly influenced by a bcc-stable point above 
the melting line.  Other points where the bcc phase was stable lie close to 
our $\widetilde{T}_{trans}$ but are shifted up due to the smaller $r_{c}$ used.

Our transition line is in qualitative agreement with more recent MD and Monte 
Carlo results \cite{HFD1997, DMRB1993}.  Dupont et.\ al.\ calculated a fcc-bcc 
coexistence point and the fcc-bcc-fluid triple point using small systems $(N 
\simeq 250)$.  Although their triple point  $(\lambda_{tp} = 6.75, 
\widetilde{T}_{tp} = 0.0034)$ lies well below ours 
$(\lambda_{tp} = 7.7 \pm 0.3, \widetilde{T}_{tp} = 0.0039 \pm 0.0001)$, it 
lies only about 2\% below our fcc-bcc transition line, and well below 
recently published melting lines \cite{SR1993,HFD1997}.  

Hamaguchi, Farouki, and Dubin also obtain a lower triple point 
$(\lambda_{tp} = 6.90, \widetilde{T}_{tp} = 0.0038)$ because their bcc-fcc 
transition temperatures are systematically (6-10\%) higher than ours 
\cite{HFD1997}.  One possible explanation is that their equilibration 
times were too short.  They used the $\lambda$-independent time unit 
$\tau = \sqrt{3}\omega_{p}^{-1}$, where $\omega_{p} = \sqrt{4\pi n\Phi/m}$ is 
the plasma frequency.  Starting with perfect bcc and fcc lattices as their 
initial conditions, they equilibrated their systems for a maximum of 300$\tau$ 
before beginning their free energy measurements.  This corresponds to about 
27$\tau_{E}$ for $\lambda = 3$ and only 4$\tau_{E}$ for $\lambda = 8$.  Since 
the latter is only about four times the velocity autocorrelation time, and 
comparable to the time for sound to propagate across their simulation cells, 
it is doubtful that their systems had equilibrated sufficiently at high 
$\lambda$.  Too short an equilibration time could cause overestimation of the 
stability of the phase with lower entropy, the fcc phase, which is consistent 
with their findings.

\begin{figure}
\includegraphics[width=3.2in]{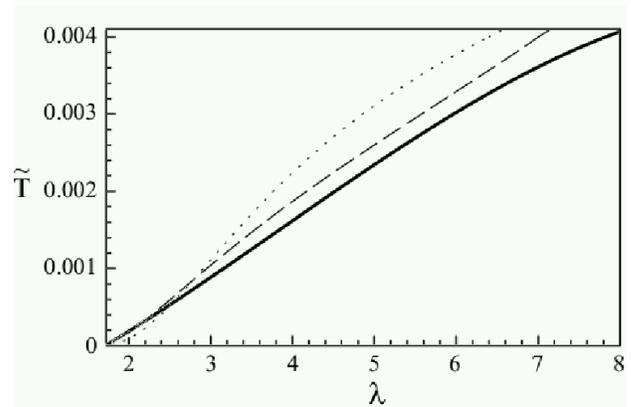}
\caption{Comparison of transition line to analytic estimates.  The heavy line 
is our $\widetilde{T}^{fit}_{trans}(\lambda)$.  The dashed and dotted lines 
are the analytic estimates for the bcc$\to$fcc and fcc$\to$bcc transitions 
from Ref.\ \protect\cite{Vaulina2002a}.}
\label{fig:comptoVaulina}
\end{figure}

Vaulina and colleagues have proposed phenomenological criteria for the bcc-fcc 
transition \cite{Vaulina2002a}.  They assume that $\kappa^{-1}$ is an 
effective hard-sphere radius and predict that the value of the rms 
displacement at the fcc transition, $\Delta_{trans}$, satisfies 
\begin{equation}
2(1 - \pi\sqrt{2}/6)^{-1/3}\Delta_{trans} = R_{WS} - \kappa^{-1}
\label{eq:geomcrit}
\end{equation}
where $R_{WS} = (4\pi/3)^{-1/3}a$ is the Wigner-Seitz radius.  They then use 
an approximate formula for the effective frequency in an Einstein-like model 
to determine $\Delta$ for the fcc and bcc structures. 
These values of $\Delta$ give two predictions for the transition line.  As 
shown in Figure \ref{fig:comptoVaulina}, their predictions are qualitatively 
correct but lie roughly 10-40\% above our $\widetilde{T}_{trans}$.  

The discrepancy in Fig.\ \ref{fig:comptoVaulina} could be due to a failure 
either of Eq.\ (\ref{eq:geomcrit}) or of the approximations used to find 
$\Delta$.  To test this we performed equilibrium simulations at 
$\widetilde{T}^{fit}_{trans}$ in both bcc $(N = 3456)$ and fcc $(N = 4000)$ 
systems.  Our results for the rms displacements, $\Delta_{fcc}$ and  
$\Delta_{bcc}$, are compared to the predictions of Eq.\ (\ref{eq:geomcrit}) in 
Table \ref{tab:rmsd}.  
The rms displacements for the fcc structure lie quite close to the prediction 
for small $\lambda$, and about 13\% above it at $\lambda = 7$.  Finite size 
effects decrease $\Delta_{fcc}$ relative to the N = $\infty$ value 
\cite{RGK1990}\footnote{Equilibrium simulations of larger (N $\simeq$ 100000) 
systems for $\lambda = 5$ showed that both $\Delta_{fcc}$ and $\Delta_{bcc}$ 
increase with N, worsening the agreement with $\Delta_{trans}$, but that the 
finite-size corrections are small compared to the discrepancies shown in Table 
VII.}  .  These results indicate that most of the error in Vaulina et.\ al.'s 
transition lines comes from substantial underestimation of the rms 
displacements.  They calculate $\Delta$ in the harmonic approximation, and 
anharmonic corrections increase $\Delta$ for these $\lambda$ \cite{RKG1988}.   
Note that the measured bcc displacements in Table VI are larger due to the bcc 
lattice's softer shear modes \cite{RKG1988}.

\begin{table}[h]
\caption{Rms displacements at $\widetilde{T}^{fit}_{trans}(\lambda)$. 
$\Delta_{trans}$ is the prediction from Eq.\ref{eq:geomcrit}, while 
$\Delta_{fcc}$ and $\Delta_{bcc}$ are results from equilibrium MD simulations.}
\begin{ruledtabular}
\begin{tabular}{llll}
$\lambda$ &  $\Delta_{trans}$ &  $\Delta_{fcc}$ & $\Delta_{bcc}$\\
3 & 0.0915 & 0.089 & 0.096\\
4 & 0.1181 & 0.118 & 0.128\\
5 & 0.1341 & 0.140 & 0.154\\
6 & 0.1447 & 0.159 & 0.177\\
7 & 0.1522 & 0.171 & 0.192\\
\end{tabular}
\end{ruledtabular}
\label{tab:rmsd}
\end{table}

\section{Summary and Conclusions}
\label{sec:conclude}
We calculated the bcc-fcc coexistence line of Yukawa systems to an uncertainty 
of approximately 1\% through integration of the mechanical work along Bain 
transformation paths.  The range of bcc stability was found to be slightly 
greater than that found in previous comprehensive studies 
\cite{RKG1988,DMRB1993,HFD1997}, and the triple point shown to lie at higher 
inverse screening length.   
The large changes in $\widetilde{T}_{trans}$ with $r_{c}$ for small $\lambda$ 
indicate that the relative stability of fcc and bcc phases depends sensitively 
on long-range correlations, and calls into question the use of local 
nearest-neighbor arguments to calculate the transition line.  Nevertheless, 
we found that one such phenomenological criterion \cite{Vaulina2002a}, 
derived from the idea that the fcc phase is stable when interparticle 
interactions are hard-sphere-like \cite{Hoover1968}, predicts the transition 
line remarkably well when combined with MD results for the mean-squared 
displacement.

Comparison with lattice-dynamics results shows that anharmonic terms in the 
total energy favor the bcc phase for all $\lambda$, but that these corrections 
are overbalanced by anharmonic contributions to the entropy.  The change in 
entropy appears to reflect an increase in the frequency of long-wavelength 
shear modes in the bcc phase.  This increase also stabilizes the bcc phase 
against a linear shear instability observed for $\lambda > 7.67$ at low 
$\widetilde{T}$. 
  
We found that shifts in the transition line due to finite size effects are 
less than 1\% if N$\sim$3000-4000, but that the presence of long-range order 
at temperatures near the transition line in weakly screened systems requires a 
cutoff radius larger than that used in some previous studies 
\cite{KRG1986, RKG1988}.  Accurate simulations of weakly screened 
($\lambda < 3$) systems in this temperature range require either larger 
system sizes or an Ewald-like summation over periodic images \cite{HFD1997}.  
However, we have also shown that a reasonably 
small potential cutoff need not introduce large errors in a transition line 
calculation in the moderate-screening regime, provided the cutoff is chosen 
with some care.

It is known that the phase behavior of real systems such as charge-stabilized 
colloidal suspensions is not fully described by pointlike Yukawa 
interactions.  Recent simulations of charged macroions in a dynamic 
neutralizing background have shown that the repulsive interactions 
between macroions are truncated by many-body effects, destabilizing 
the crystalline phases in the weak screening limit
\cite{Dobnikar2003a,Dobnikar2003b,Dijkstra2003b}.  Independently, hard-core 
repulsions significantly alter the phase diagram when the volume fraction is 
more than a few percent \cite{ME1997,EBRM2000,Dijkstra}.    However, we hope 
that our high-precision calculation of the point Yukawa fcc-bcc transition 
line may serve as a benchmark for further studies of more sophisticated models.

\section{Acknowledgements}
The simulations in this paper were carried out using the LAMMPS molecular 
dynamics software.  Support from NSF Grant DMR-0083286 is gratefully 
acknowledged.
 

\end{document}